\documentstyle[aps,epsf,twocolumn]{revtex}
\begin{document}
\draft
\title{Exact calculation of thermodynamical quantities of the 
       integrable $t-J$ model}
\author{G.~J\"uttner and A. Kl\"umper}
\address{Universit\"at zu K\"oln  \\                
        Institut f\"ur Theoretische Physik \\     
        Z\"ulpicher Str. 77  \\                   
        D-50937, Germany}
\date{January, 1996}
\maketitle
\begin{abstract}
The specific heat and the compressibility for the
integrable $t-J$ model are calculated showing Luttinger liquid
behavior for low temperatures.
A Trotter-Suzuki mapping and the quantum transfer matrix approach are
utilized. Using an algebraic Bethe ansatz this method permits the
exact calculation of the free energy and related quantities. A set
of just two non-linear integral equations determining these
quantities is studied for various particle densities and
temperatures. The structure of the specific heat is discussed in
terms of the elementary charge as well as spin excitations.
\end{abstract}

The $t-J$ model has  been studied intensively  in recent years because
of its importance  as one of  the most fundamental systems of strongly
correlated electrons and its possible  relevance for purely electronic
mechanisms for high-$T_c$ superconductivity \cite{And87,ZhaRi88}.  The
model  describes   the   nearest-neighbor hopping   of  electrons with
spin-exchange interaction.  The  effect of a strong  repulsive on-site
Coulomb interaction is modeled by the restriction of the Hilbert space
to states without doubly  occupied lattice sites.  The one-dimensional
Hamiltonian reads 
\begin{eqnarray}
  H&=&-t\sum_{j,\sigma}{\cal P}
                     (c^\dagger_{j,\sigma}c_{j+1,\sigma}+
                      c^\dagger_{j+1,\sigma}c_{j,\sigma})
                     {\cal P}\nonumber\\
    &&+J\sum_j(S_j S_{j+1}-n_j n_{j+1}/4)
  \label{hamilton}  
\end{eqnarray}
where the projector ${\cal P}=\prod_j(1-n_{j\uparrow}n_{j\downarrow})$
ensures that double   occupancies  of sites  are  forbidden.  At   the
supersymmetric  point $2t=J$ the  system was  shown  to be  integrable
\cite{Suth75,Sch87} by  the  well-known Bethe ansatz. The
ground state and excitation spectrum were investigated by Bares et
al. \cite{BarBlaOg91}. Critical exponents were  calculated by finite-size
scaling and conformal  field theory \cite{KawaYa90,BarKlu95}. 

According to  the    pioneering  work   of   Takahashi    \cite{Tak71}
thermodynamical properties of general integrable systems are described
by  an infinite  set of  coupled   integral equations reflecting   the
existence of   infinitely many different   rapidity patterns.  For the
$t-J$ model such integral  equations were formulated in 
\cite{Will92,Sch92},
but  not solved.   We  like  to point   out  two shortcomings  of  the
traditional  approach. First, the method  consists  of a rather direct
evaluation  of the   partition function by   taking  into account  all
excited states of the Hamiltonian. The excitations  are derived on the
basis of the so-called string conjecture which describes the solutions
of  the    Bethe ansatz equations. However, the validity of the string
conjecture is quite controversial. Second, the calculation of the 
asymptotics of correlation functions is not possible within the
traditional approach. 

Our approach takes a  different route following \cite{Suz85,SuIn87} to
overcome the mentioned problems. We do not  directly study the Hamiltonian 
at finite temperatures. Instead, we employ a convenient
Trotter-Suzuki mapping \cite{Suz85} to an
exactly solvable two-dimensional classical model. Furthermore, the
corresponding thermodynamical quantities  are expressed by eigenvalues
of the so-called quantum transfer  matrix. The largest
eigenvalue,  for example,   directly yields   the free  energy.    The
next-largest   eigenvalues  provide   the   correlation lengths.
This  represents the important  advantage  compared with the
traditional thermodynamical  Bethe ansatz requiring all eigenvalues of
the Hamiltonian.  The  new method using  the Trotter-Suzuki mapping as
well  as  the quantum  transfer  matrix approach has  been  applied to
several   quantum systems \cite{Suz85,SuIn87,Suz90,Tak91,Klu93,DeVe92}
notably the Hubbard model in \cite{KluBa95}. 

The $t-J$ quantum chain  belongs to a  family of integrable  fermionic
systems which   are related   to   the classical   Perk-Schulz   model
\cite{Sch81},   a  multi-component generalization   of  the well-known
six-vertex  model.  The detailed  derivation  of the quantum  transfer
matrix and the study  of the corresponding  eigenvalue problem  by the
quantum  inverse scattering  method  will be  described  in a separate
publication \cite{KluWe96,JuKlu96}.  Thus, we confine ourselves to the
main points. 

The  Perk-Schultz model is exactly solvable  as  the Boltzmann weights
satisfy the Yang-Baxter equation. This in turn implies the commutation
of all row-to-row transfer  matrices for arbitrary spectral parameters
$u$, $v$: $T(u)T(v)=T(v)T(u)$.  Consequently, the Hamiltonian is
integrable. We have
\begin{equation}
  H=\frac{d}{du}\ln T(u)_{|_{u=0}}\quad\mbox{with}\quad
  T(u)=T_R\, e^{uH+{\cal O}(u^2)},
\label{qtm-exp}
\end{equation}
where $T_R$ is a  right-shift operator.  By  means of the substitution
$u=-\beta/N$ ($\beta$ being the inverse temperature) we find 
\begin{equation}
  \big(T(-\beta/N)\overline{T}(-\beta/N)\big)^{N/2}=e^{-\beta H+{\cal
      O}(1/N)},
\label{qtm-element}
\end{equation}
where   $\overline{T}$  is   the  row-to-row transfer    matrix of the
Perk-Schultz   model    after  a   $90^o$   rotation.    It  satisfies
(\ref{qtm-exp}) with  $T,\ T_R$ replaced by $\overline{T},\ T_R^{-1}$.
The partition function of the quantum system
\begin{equation}
  Z=\lim_{N\to\infty}{\rm Trace}
  \big(T(-\beta/N)\overline{T}(-\beta/N)\big)^{N/2}
\end{equation}
is identical   to  the partition     function of  the    inhomogeneous
Perk-Schulz  model    with  alternating rows   \cite{KluWe96}.     The
technically more convenient column-to-column transfer matrix of such a
system  is often referred  to as  the  quantum transfer matrix  (QTM).
%% New:
This matrix   is  shown  to  be  embedded   into a   commuting  family
\cite{KluWe96,JuKlu96}  of   staggered    transfer matrices given   by
alternating products of unrotated and rotated  vertices. We still have
the  same intertwiner as in the  case of the  homogeneous lattice.  In
contrast to  the work in  \cite{KluBa95}  for  the Hubbard  model  the
integrability is preserved  within  the present  approach.   Thus, the
eigenvalues of   the QTM  are   analytic  functions  of  the  spectral
parameter.  This fact represents   the important advantage  leading to
simple calculations which can be generalized to other models. 
The free energy can be  calculated by the largest eigenvalue $\Lambda$
of  the  QTM  \mbox{${f}=-(\log\Lambda)/\beta$} and  the  next-leading
eigenvalues yield the correlation lengths.

We   treat the quantum  transfer matrix  by   using an algebraic Bethe
ansatz  \cite{KluWe96}.  The  eigenvalues $\Lambda(v)$   of $T(v)$ are
shown to be analytic functions of the spectral parameter $v$ with 
explicit representation 
\begin{equation}
  \Lambda(v)=\lambda_1(v)+\lambda_2(v)+\lambda_3(v),
  \label{evlambda}  
\end{equation}
where
\begin{eqnarray*}
  \lambda_1(v)&=&\prod_{j}\frac{v-w_j+i}{v-w_j}
               \Big[(v-i u-i)(v+i u)\Big]^{N/2}
               e^{\beta\mu},
  \\
  \lambda_2(v)&=&\prod_{k}\frac{v-v_k-i}{v-v_k}
              \Big[(v+i u+i)(v-i u)\Big]^{N/2}
              e^{\beta\mu},
  \\
  \lambda_3(v)&=&\prod_{j}\frac{v-w_j+i}{v-w_j}
              \, (v+i u)^{N/2}\times
  \\
              &&\prod_{k}\frac{v-v_k-i}{v-v_k}
              \, (v-i u)^{N/2}
\end{eqnarray*}
and $v_j$  and $w_j$  are  Bethe ansatz rapidities,  $\mu$ denotes the
chemical potential measured from the lower edge of the band.  
The defining relations for the rapidities, the so-called Bethe ansatz 
equations are equivalent to the analyticity of $\Lambda(v)$, i.e. the 
absence of poles. We use this  analyticity to determine the largest
eigenvalue by a finite set of
non-linear  integral  equations  \cite{KluWe96,JuKlu96}.   This   also
allows for taking   the limit $N\to\infty$ analytically.  
%% New:
We define the following auxiliary  functions (with arbitrary but fixed
\mbox{$0<\gamma<1$}) 
\begin{eqnarray*}
  b(x)&=&\frac{\lambda_1(x+i\gamma)}{\lambda_2(x+i\gamma)+ 
                                   \lambda_3(x+i\gamma)},\\
  c(x)&=&\frac{\lambda_1(x)\lambda_2(x)}{\lambda_3(x)
             \big(\lambda_1(x)+\lambda_2(x)+\lambda_3(x)\big)}.
\end{eqnarray*}
Using this  definition and the  analyticity property of the eigenvalue
$\Lambda(v)$ a set of functional equations for $b$ and $c$ is derived.
This  set can  be   transformed into  integral equations  by  standard
applications of   the Fourier transform.   Such  an approach should be
applicable to all integrable models.
After some lengthy calculations \cite{JuKlu96} we obtain the following
relations for the two auxiliary functions 
\begin{eqnarray*}
  \log b(x)&=&-2\pi\beta\Psi_b(x+i\gamma)+\beta\mu-
  \\
  &&\Psi_b\ast\log(1+\overline{b})|_{{x+2i\gamma}}
  -\Psi_b\ast\log(1+c)|_{{x+i\gamma}},
  \\
  \log c(x)&=&-2\pi\beta\Psi_c(x)+2\beta\mu-
  \\
  &&2\Re\big(\Psi_b\ast\log(1+\overline{b})|_{{x+i\gamma}}\big)
  -\Psi_c\ast\log(1+c)|_{{x}},
\end{eqnarray*}
with
\begin{displaymath}
    \Psi_b(x)=1/\big(2\pi x(x-i)\big),\quad 
    \Psi_c(x)=2/\big(2\pi (x^2+1)\big),
\end{displaymath}
where $\ast$ denotes the convolution $f\ast g=\int f(x-y)g(y)dy$ taken
at   the    indicated   arguments   $x$,     \mbox{$x+i\gamma$}    and
\mbox{$x+2i\gamma$}. The solution  of the integral equations  provides
the largest eigenvalue via: 
\begin{equation}
  \log\Lambda=-\log c(0)+2\beta\mu.
\end{equation}
In order to calculate derivatives of the thermodynamical potential one
can  avoid  numerical differentiations  by utilizing  similar integral
equations \cite{JuKlu96} guaranteeing  the same  numerical accuracy as
for of the free energy. 

In the  low-temperature and low-density  limit 
(small $\mu$ and $T/\mu$) we obtain an analytical
expression       for      the        grand      canonical    potential
\mbox{${f}=-(\log\Lambda)/\beta$}
\begin{equation}
f=-\int\limits_{-\infty}^\infty\frac{dx}{\pi\beta^{3/2}}\, 
  \log\big(1+e^{-x^2+\beta\mu}\big), 
\end{equation}
in agreement with the numerical results. Asymptotically we find
\begin{equation}
  {f}=-\frac{4}{3\pi}\mu^{3/2}-
  \frac{\pi}{6\sqrt{\mu}}\, T^2+{o}(\mu^{3/2},T^2/\sqrt{\mu}),
\label{pot}
\end{equation}
implying a particle density \mbox{$n=-\partial{f}/\partial\mu$} and
entropy \mbox{$S=-\partial{f}/\partial{T}$}: 
\begin{equation}
  n=\frac{2}{\pi}\sqrt{\mu},\quad S=\frac{2}{3n}\, T.
\label{entropy}
\end{equation}
Moreover, due to predictions of Luttinger  liquid theory and conformal
field theory we expect for the low-temperature asymptotics 
\begin{equation}
  f=f_0-\left(\frac{\pi c_s}{6 v_s}+\frac{\pi c_c}{6 v_c}\right)T^2
\label{cft}
\end{equation}
where $v_{s,c}$  and $c_{s,c}$ are the  velocities and central charges
(\mbox{$c_{s,c}=1$}) for the  elementary spin  and charge excitations.
For   small particle    densities   we  have   \mbox{$v_{s,c}=\pi{n}$}
\cite{BarBlaOg91}, thus (\ref{pot}) and (\ref{cft}) are consistent. 

On the other hand, the simple  high-temperature limit for finite $\mu$
immediately  leads  to \mbox{$S=\ln{3}$}   (and \mbox{$n=2/3$})  as is
expected by  counting  the  degrees   of freedom  per   lattice  site.
Therefore, the  analytical   benchmarks   for  \mbox{$T\to 0$}     and
\mbox{$T\to\infty$} imply the consistency of our approach.

\begin{figure}[htbp]
  \begin{center}
    \leavevmode
    \epsfxsize=0.35\textwidth\epsfbox{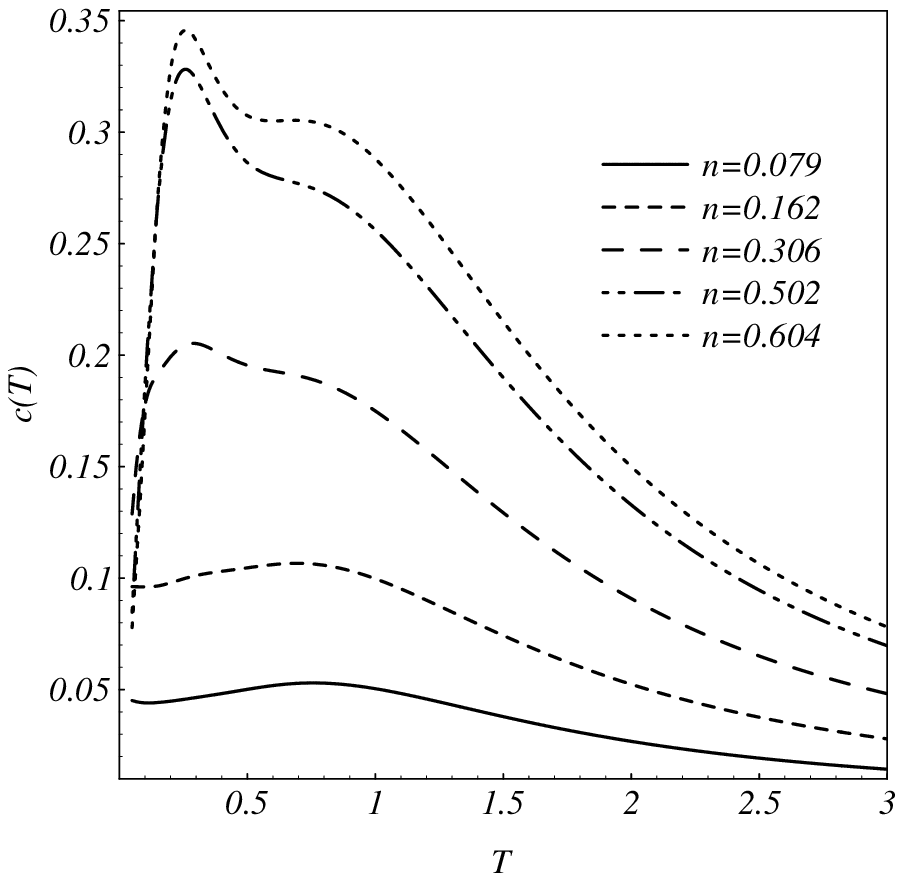}
    \epsfxsize=0.35\textwidth\epsfbox{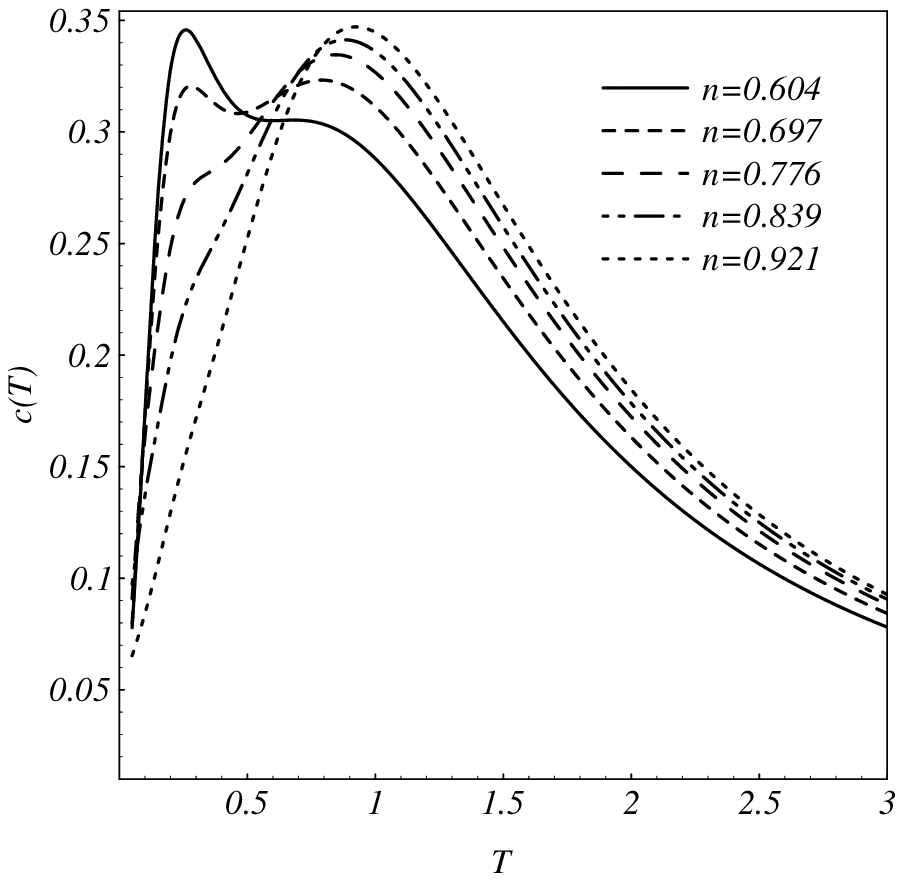} 
    \caption{Specific heat as function of $T$ for different 
      particle densities $n$ with $n\le 0.6$ and $n\ge 0.6$.}
    \label{fig1ab}
  \end{center}
\end{figure}

\begin{figure}[htbp]
  \begin{center}
    \leavevmode
    \epsfxsize=0.35\textwidth\epsfbox{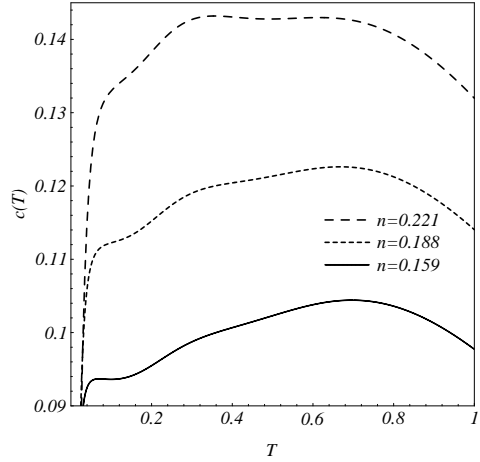} 
    \caption{Specific heat at low temperatures $T$ for intermediate 
      particle densities $n$.}
    \label{fig2}
  \end{center}
\end{figure}

Now we  consider  various  thermodynamical quantities  at intermediate
temperatures   by  solving    the    integral equations   numerically.
Figure~\ref{fig1ab} shows the specific  heat
as a function of $T$ for  different fixed particle densities. First of
all, we note a linear temperature  dependence at low $T$. According to
conformal  field theory  (\ref{cft})   the coefficient   is given   by
\mbox{$\pi({1/v_s}+{1/v_c})/3$}.  Our   numerical data  are consistent
with this expression.
Furthermore,   we  observe two  maxima   with  changing dominance  for
increasing particle density $n$.  The  nature of this structure can be
understood  from  the  elementary excitations  of  the  system. In the
groundstate the particles  are  bound in  singlet  pairs with  binding
energies varying from 0 to some density dependent value. There are two
types of excitations.   First,  there are  charge excitations  due  to
energy-momentum transfer  onto  individual  pairs. Second, there   are
excitations due to the breaking of pairs.  The latter excitation is of
spin type at  lower excitation energies, but  changes the character at
higher (density dependent) energies to charge type as it describes the
motion of single particles. Therefore, the first and second maximum at
lower densities (figure \ref{fig1ab}.a) are caused by charge
excitations due  to  pairs  and  single  particles,  respectively.  At
higher densities (figure \ref{fig1ab}.b) the maximum at lower
temperatures is dominated by excitations  of pairs whereas the  second
one  at  higher   temperatures is  caused  by  spin  excitations.  For
increasing concentration the spin contribution becomes dominant as the
charge   excitations freeze out.   This  is   in accordance with   the
limiting case $n=1$ leading  to  the spin-1/2  Heisenberg  chain.  The
missing spin  structure in the specific  heat at  low and intermediate
densities  is    found    at   quite  low     temperatures   shown  in
figure~\ref{fig2}. 

It  is worthwhile to compare these  results  with the findings for the
Hubbard model  investigated by the  traditional thermodynamical  Bethe
ansatz \cite{UsuKawa90}.  The structure found in  the specific heat is
explained by spin and  charge  excitations which  do not  change their
character in  contrast to the $t-J$  model.   For certain  densities a
low-temperature  charge peak  was found   which  is caused by   single
particle  excitations.  A charge   peak at higher temperatures appears
because  of excitations due  to  doubly  occupied lattice  sites  with
energies of the order $U$ for large Coulomb interaction. 

\begin{figure}[htbp]
  \begin{center}
    \leavevmode 
    \epsfxsize=0.35\textwidth\epsfbox{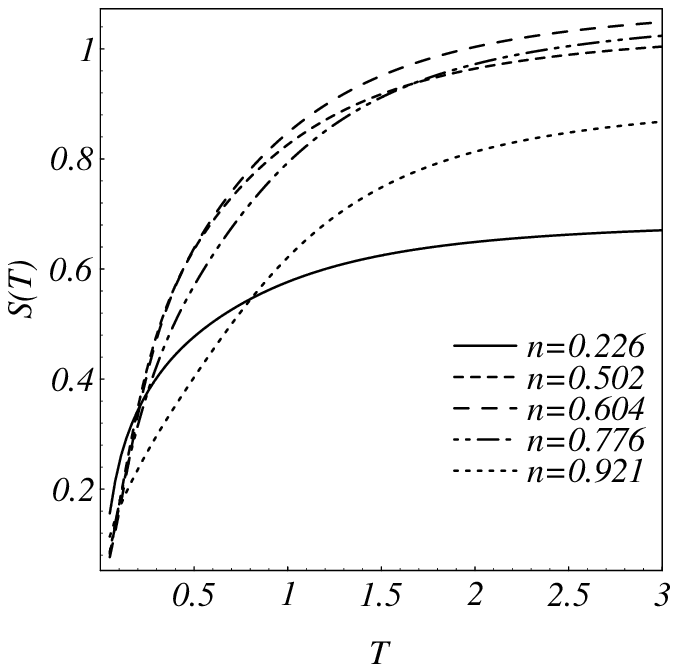}
    \epsfxsize=0.35\textwidth\epsfbox{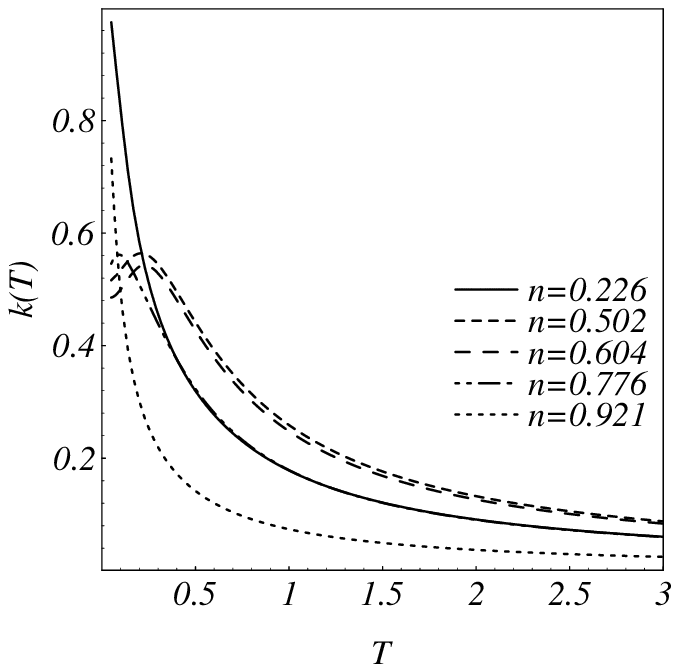}
    \caption{Entropy $S$ and compressibility $\kappa$ 
      versus $T$ for different densities $n$.}
    \label{fig3ab}
  \end{center}
\end{figure}

To  conclude our investigation we present  numerical results for other
thermodynamical quantities.  Figure~\ref{fig3ab}  presents the
entropy               and             the              compressibility
\mbox{$\kappa=\partial{}n/\partial\mu$}     for     various   particle
densities.   Note the  divergent  low-temperature compressibility  for
particle densities \mbox{$n\to{0}$} and \mbox{$n\to{1}$}. 

In summary,   we have  derived eigenvalue  equations  for the  quantum
transfer matrix  of the $t-J$ model.  This  approach permits  the {\it
  exact}    analytical  as   well   as    numerical   calculation   of
thermodynamical quantities.   Instead of  solving an  infinite  set of
integral equations     --   as is     necessary  in  the   traditional
thermodynamical Bethe ansatz -- we have to solve integral equations for
only {\it two} functions. We have considered  analytically certain low
and high-temperature limits verifying the values predicted by conformal
field theory.  Moreover,   the case of intermediate   temperatures was
treated  numerically.   As   shown,  the  specific  heat 
and compressibility display   an
interesting    behavior   in dependence   on   particle   density  and
temperature. 

The authors acknowledge financial support by the 
{\it Deutsche Forschungsgemeinschaft} under grant No. Kl~645/3-1.

\onecolumn

\noindent
\noindent

\end{document}